\documentclass[acmsmall]{acmart}

\usepackage{setspace}
\usepackage{graphicx}
\usepackage{subcaption}
\usepackage{soul, xcolor}

\AtBeginDocument{
  \providecommand\BibTeX{{
    \normalfont B\kern-0.5em{\scshape i\kern-0.25em b}\kern-0.8em\TeX}}}

\setcopyright{acmcopyright}
\copyrightyear{2021}
\acmYear{2021}
\acmDOI{}

\acmJournal{JETC}
\acmVolume{Special issue on Trustworthy AI}

\begin{document}

\title{Power-Based Attacks on Spatial DNN Accelerators}

\author{Ge Li}
\email{lige@utexas.edu}
\orcid{0000-0003-1075-6859}
\author{Mohit Tiwari}
\email{tiwari@utexas.edu}
\orcid{0000-0002-0384-3308}
\author{Michael Orshansky}
\email{orshansky@utexas.edu}
\orcid{0000-0002-6223-4748}
\affiliation{
  \institution{The University of Texas at Austin}
  \department{Department of Electrical and Computer Engineering}
  \streetaddress{2501 Speedway}
  \city{Austin}
  \state{Texas}
  \country{USA}
  \postcode{78712}
}

\begin{abstract}
With proliferation of DNN-based applications, the confidentiality of DNN model is an important commercial goal. Spatial accelerators, that parallelize matrix/vector operations, are utilized for enhancing energy efficiency of DNN computation. Recently, model extraction attacks on simple accelerators, either with a single processing element or running a binarized network, were demonstrated using the methodology derived from differential power analysis (DPA) attack on cryptographic devices. This paper investigates the vulnerability of realistic spatial accelerators using general, 8-bit, number representation.

We investigate two systolic array architectures with weight-stationary dataflow: (1) a 3 $\times$ 1 array for a dot-product operation, and (2) a 3 $\times$ 3 array for matrix-vector multiplication. Both are implemented on the SAKURA-G FPGA board. We show that both architectures are ultimately vulnerable. A conventional DPA succeeds fully on the 1D array, requiring 20K power measurements. However, the 2D array exhibits higher security even with 460K traces. We show that this is because the 2D array intrinsically entails multiple MACs simultaneously dependent on the same input. However, we find that a novel template-based DPA with multiple profiling phases is able to fully break the 2D array with only 40K traces. Corresponding countermeasures need to be investigated for spatial DNN accelerators.

\end{abstract}



\maketitle

\section{Introduction}
Progress in Deep Neural Networks (DNNs) has been driving various applications including computer vision \cite{resnet}, \cite{mobilenet}, speech recognition \cite{cnn_speech_recognition}, \cite{dnn_speech_recognition}, medical image analysis \cite{dl_medical_1}, \cite{dl_medical_2}, malware detection \cite{malware_dnn_1}\cite{malware_dnn_2} and many others. DNNs enable learning complex features in input data \cite{efficient_dnn_survey}, achieving superior performance in a variety of tasks, compared to conventional machine learning algorithms.

Excellent performance of DNNs depends on tremendous effort to train the neural network model. The cost of model creation can be captured in the following aspects: (1) Labor and time to create/collect a dataset for training, (2) the cost of computing resources to run the DNN training algorithm, and (3) the search for hyper-parameters which result in optimal DNN models. Therefore, DNN models for specific applications need to be considered a form of intellectual property with high commercial value. This creates a motivation to obtain the DNN models via adversarial/non-commercial means, e.g. via attacks that extract the model. Besides the loss of commercial value, the loss of DNN model confidentiality can also lead to security and privacy problems. An exposed model may facilitate adversarial attacks, where an attacker crafts input samples that make the target model mis-classify \cite{adversarial_attack}, or, membership inference attacks, in which an attacker aims to learn whether an input to the target model belongs to its private training set \cite{membership_infer_attack}. 

 Model extraction attacks utilizing class probabilities (confidence scores) from model output have been demonstrated. In \cite{ml_model_extract_api}, Tram\`{e}r et al., demonstrated a model extraction attack against some DNNs by observing outputs of API queries. In \cite{high_fidelity_extraction},  Jagielski et al., showed an efficient learning-based model extraction attack that utilizes query outputs. Query-based attacks can be effectively defended by limiting model output. Side-channel attacks have also been demonstrated as effective for DNN model extraction \cite{dnn_arch_power, masked_net, csi_nn, cema_single_pe, mem_access_cnn_attack, cache_dnn_attack, neural_net_extraction_arch, timing_neural_net_attack}. Side-channel attacks offer a significant risk because side-channel information emanating during DNN execution is difficult to eliminate. Exploitation of digital side channels enables the attacker to extract coarse-grained information of the target DNN model, such as its architecture. In \cite{mem_access_cnn_attack}, Hua et al., proposed a reverse-engineering attack based on observing the accelerator off-chip memory access patterns and enumerating the possible architectures that satisfy the constraints. In \cite{cache_dnn_attack}, Yan et al., utilized both memory access and timing side channels to reverse-engineer the size of matrices involved in matrix multiplication allowing to infer the DNN architecture. Hu et al., extracted the DNN architecture from the noisy PCIe and memory-bus events on a GPU platform \cite{neural_net_extraction_arch}. Duddu et al. utilized execution time to infer target DNN layer depth \cite{timing_neural_net_attack}. 

As this paper demonstrates, attacks that exploit device power consumption or EM emanation enable a direct retrieval of DNN weights. In an attack modality, borrowed from attacks on cryptographic implementations, the attacker feeds a large number of inputs and collects corresponding power/EM traces using a hypothesis-testing foundation. The attacker selects a power model which represents the power of an intermediate state of the secret-related computation, makes a hypothesis on the secret, and evaluates the power model based on the hypothesis. The correct hypothesis results in the highest correlation between power/EM traces and predicted power values. 

Several efforts have demonstrated such attacks for DNN weight retrieval. In \cite{csi_nn}, Batina et al., demonstrated a correlation EM attack on a micro-controller to retrieve approximate values of single-precision weights of a multi-layer perceptron. While the attack is demonstrated on a conventional micro-controller, its feasibility on a customized DNN hardware accelerator with high parallelism is unexplored. In \cite{cema_single_pe}, Yoshida et al., implemented an FPGA-based DNN accelerator with a single processing element (PE) and performed a correlation EM attack to retrieve the weights, stored in a 8-bit fixed-point representation, by analyzing the multiply-and-accumulate operation. However, this work has not studied the feasibility of an attack on a high-performance accelerator with multiple processing elements, which is a more common deployment scenario. Dubey et al. demonstrated a differential power attack (DPA) to retrieve the binarized weights of a model implemented in an FPGA-based DNN accelerator with an adder tree used for accumulation \cite{masked_net}. The attack is only evaluated against binarized weights. Whether the attack can be effective against a 8-bit weight implementation, which is typically used by state-of-the-art accelerators, needs to be examined.

In this work, we demonstrate a differential power attack to retrieve the weights from the FPGA-based matrix multiplication accelerator. We adopt a 8-bit integer (INT8) weight representation and implement a design with multiple processing elements performing parallel multiply-and-accumulate (MAC) operations. The design adopts a weight-stationary dataflow. Matrix multiplication between input activations and weights is the core computational kernel of DNNs, as computation of both convolutional and fully-connected layers can be mapped to matrix multiplication. We propose a multi-step DPA framework which utilizes the dependency between MAC results of different weights to determine the value of each weight using measured power traces. 

We study both 1D and 2D systolic arrays. We consider a 1D array as a separate case because it is an important VLSI model. In addition, compared to 2D arrays, it can offer better reconfigurability, lower memory bandwidth requirement, and better energy efficiency due to the reduced inter-PE communication and control logic complexity, which may be desired in certain scenarios \cite{Kung_1D_array, chain_nn}.  We first study a 3 $\times$ 1 systolic array. The results show that we are able to retrieve all weights in the weight vector of the 1D array using 20K power traces with a conventional DPA. Next, we study the scalability of the DPA attack to larger designs. We implement a 3 $\times$ 3 systolic array.
The results show that the 2D accelerator exhibits higher security. Both a  conventional DPA and a stronger, template-based DPA fail: only 30\% of weights are recovered after 460K traces. We investigate the causes of higher resistance of the 2D array to the attack compared to the 1D accelerator. We explain the reason for higher security of a 2D array design by the fact that it intrinsically entails multiple MAC outputs that are simultaneously dependent on the same input. We show that this is fundamentally a feature of the weight-stationary dataflow. However, we discover that an enhanced template-based DPA with multiple profiling phases manages to expose leakage of each PE column step-by-step and retrieves weights from each column. The attack is able to retrieve all weights from the 2D array with only 40K traces. The results on both 1D and 2D arrays show that both architectures are ultimately vulnerable.

Our contribution can be summarized as follows: 

\begin{itemize}
    \item We investigate the vulnerability of a 1D systolic array. Our results show that a conventional DPA on the 1D array succeeds fully, requiring 20K power measurements.
    \item We investigate the vulnerability of a 2D systolic array. However, only 30$\%$ of the model weights are retrieved even after 460K power traces with a conventional DPA and a stronger, template-based DPA. Our analysis finds that the higher security of a 2D array arises from the fact that it intrinsically entails multiple MACs that are simultaneously dependent on the same input.
    \item We propose a novel template-based DPA with multiple profiling phases which fully breaks the 2D systolic array.
    \item We investigate Hamming distance, Hamming weight, and bit-level power models in the attack on a 2D array. 
\end{itemize}

\section{DNN Accelerator Design}
\subsection{DNN and Matrix Multiplication}
The computation of major DNN layers, including fully-connected layers and convolution layers, can be mapped to matrix multiplication. The fully-connected layer computes a weighted sum of all its input activations for each output activation. For a fully-connected layer with $M$ input neurons and $N$ output neurons, this process can be summarized as the multiplication between an $N \times M$ weight matrix and an $M \times 1$ input activation vector, where each row of the weight matrix represents the weights used to calculate the corresponding output activation. The convolution layer computes a dot product between the filter weights and input feature map pixels within the filter for each output feature map pixel. The output feature map is computed by sliding the filter with a certain stride. This computation process can be mapped to matrix multiplication between the filter weights and the input activations as well. For a convolution layer with $C_{in}$ input channels, $C_{out}$ output channels, $W_{in} \times H_{in}$ input feature map, $W_{f} \times H_{f}$ filter, stride $S$, and padding size $P$, the filter weights can be represented by a matrix with $C_{out}$ rows and $C_{in} \times W_{f} \times H_{f}$ columns. The input feature map pixels can be organized as a matrix with $C_{in} \times W_{f} \times H_{f}$ rows and $W_{out} \times H_{out}$ columns, where $W_{out}$ and $H_{out}$ represent width and height of the output feature map and can be computed as:

\begin{equation}
    W_{out} = (W_{in}-W_{f} + 2P)/S + 1
\end{equation}

\begin{equation}
    H_{out} = (H_{in}-H_{f} + 2P)/S + 1
\end{equation}

We now provide the details of the implementation of the systolic array matrix multiplication accelerator for neural networks. We implemented two versions of the accelerator: a 1D systolic array with 3 PEs, and a 2D systolic array with 9 PEs. Each PE of the accelerator can perform a 8-bit MAC on signed integers. The systolic array uses a weight-stationary dataflow, similar to Google's TPU design \cite{tpu}. Since the 1D systolic array computes the dot product while the 2D systolic array computes the matrix-vector multiplication over the input vector, we refer to the 1D systolic array design as the dot product accelerator and to the 2D systolic array design as the matrix-vector multiplication (MVM) accelerator.

\subsection{Dot Product Accelerator}

\begin{figure}[t!]
  \includegraphics[width=0.75\linewidth]{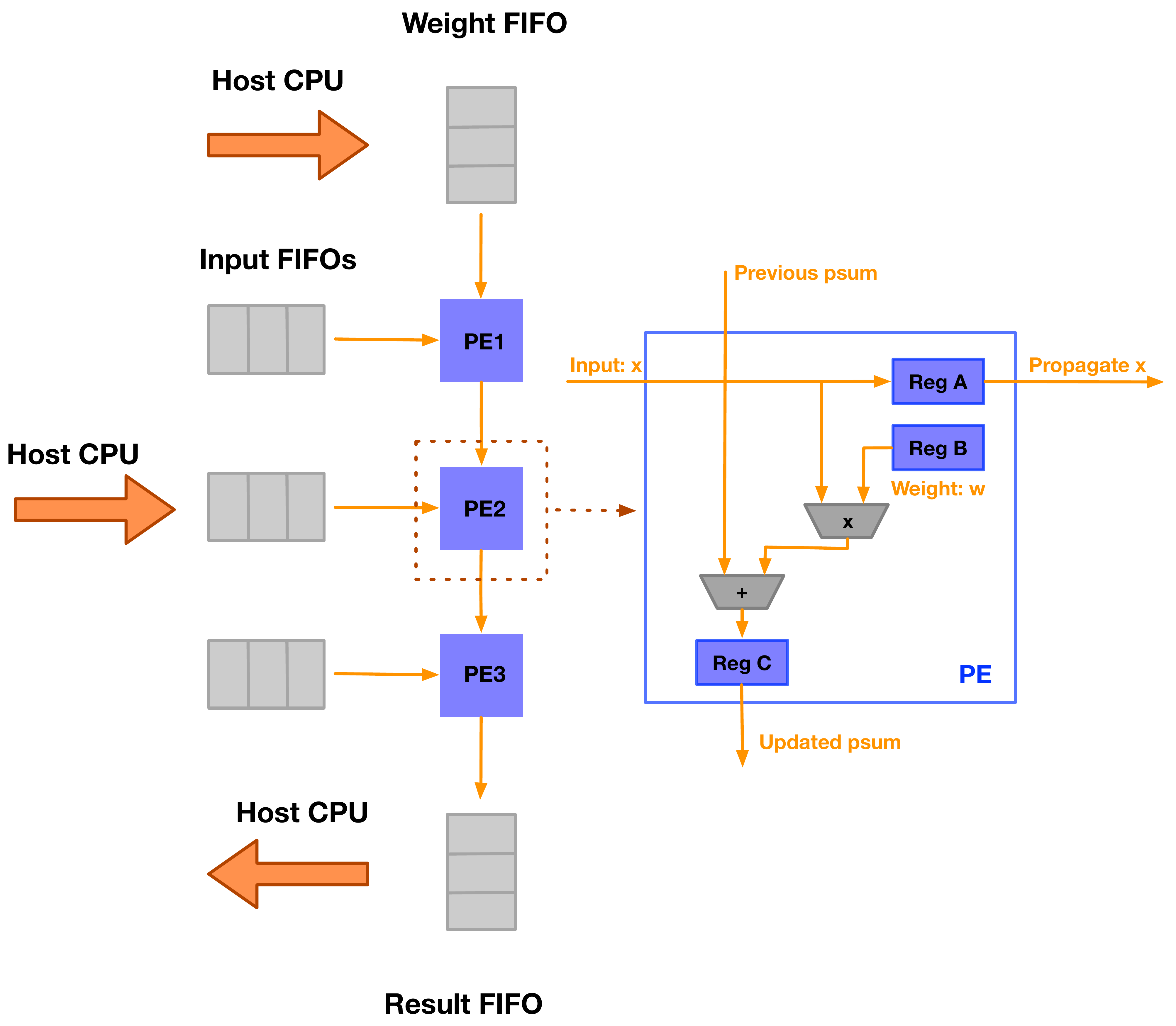}
  \caption{A 1D dot product array that uses weight-stationary dataflow.}
  \label{dp_accel}
\end{figure}

The dot product array consists of input, weight, and result FIFOs and 3 PEs arranged in a single column, as shown in Fig. \ref{dp_accel}. Each PE contains a multiplier and an adder, as well as registers to hold the input, the weight and the resulting partial sum. First, the accelerator receives inputs and weights from a host CPU and pushes them into the FIFOs. The design has the input FIFO depth of 3, which means that the dot products for 3 3\(\times\)1 input vectors are computed for each host-to-accelerator data transfer. Next, the weights are popped into a register of the corresponding PE. After pre-loading all weights into PEs, the main phase of the dot product computation begins.

\begin{figure}[t!]
  \includegraphics[width=0.75\linewidth]{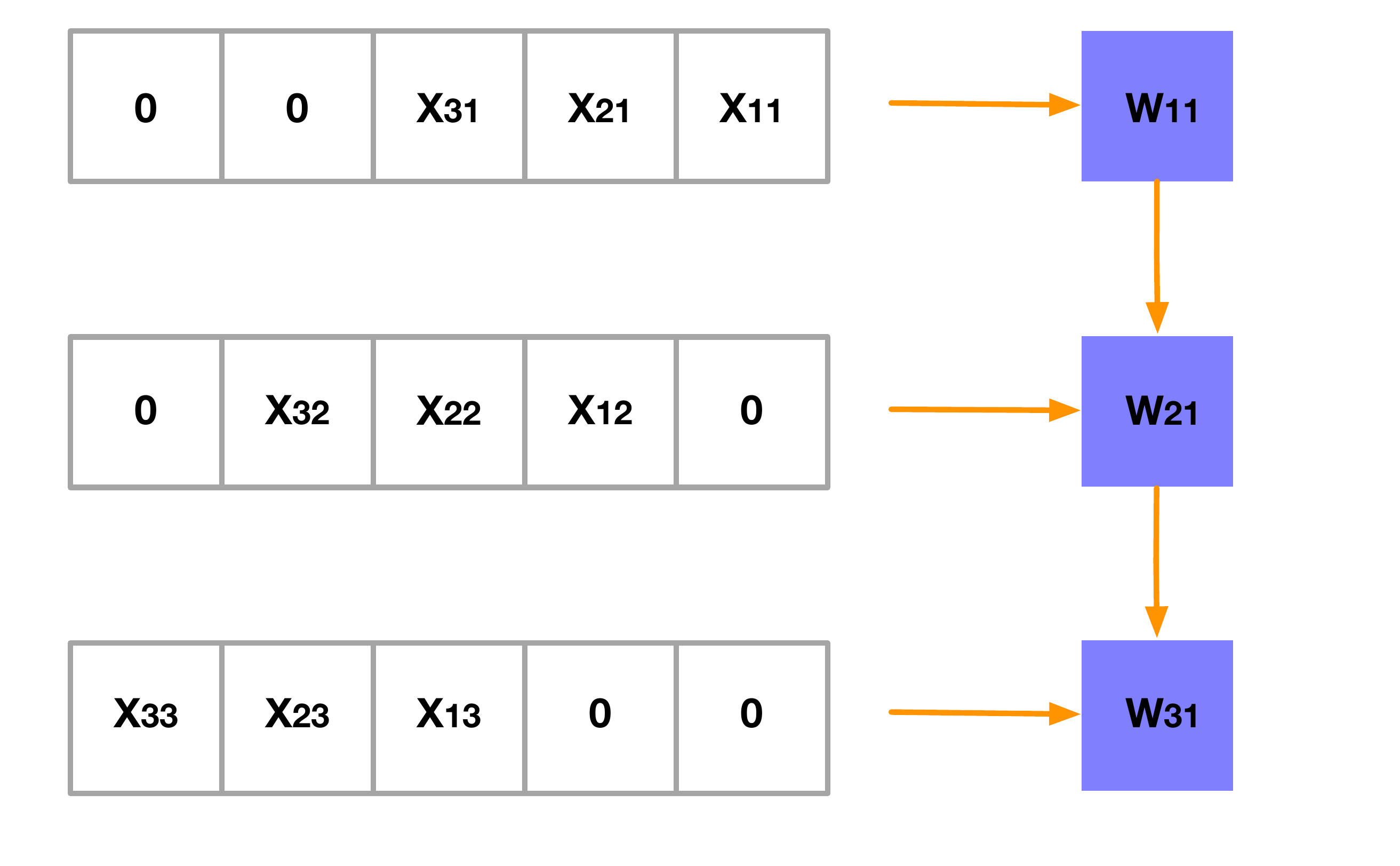}
  \caption{The diagonal wave-front propagation of inputs in the dot product accelerator.}
  \label{input_dp_accel}
\end{figure}

During the computation phase, the inputs are propagated from input FIFOs to PEs. The propagation of inputs is arranged in a diagonal wave-front format: the start of data propagation for adjacent rows of input FIFOs differs by one clock cycle, as shown in Fig. \ref{input_dp_accel}. Each clock cycle the inputs multiply the weight in each PE. The result is accumulated with the partial sum from a previous PE. The updated partial sum is propagated down to the next PE. The first PE performs only multiplication as there is no previous partial sum. It takes 5 clock cycles to compute the dot products for 3 input vectors. The results are pushed into the result FIFO, and read out by the host machine.

\subsection{Matrix-Vector Multiplication Accelerator}

\begin{figure}[t!]
  \includegraphics[width=0.7\linewidth]{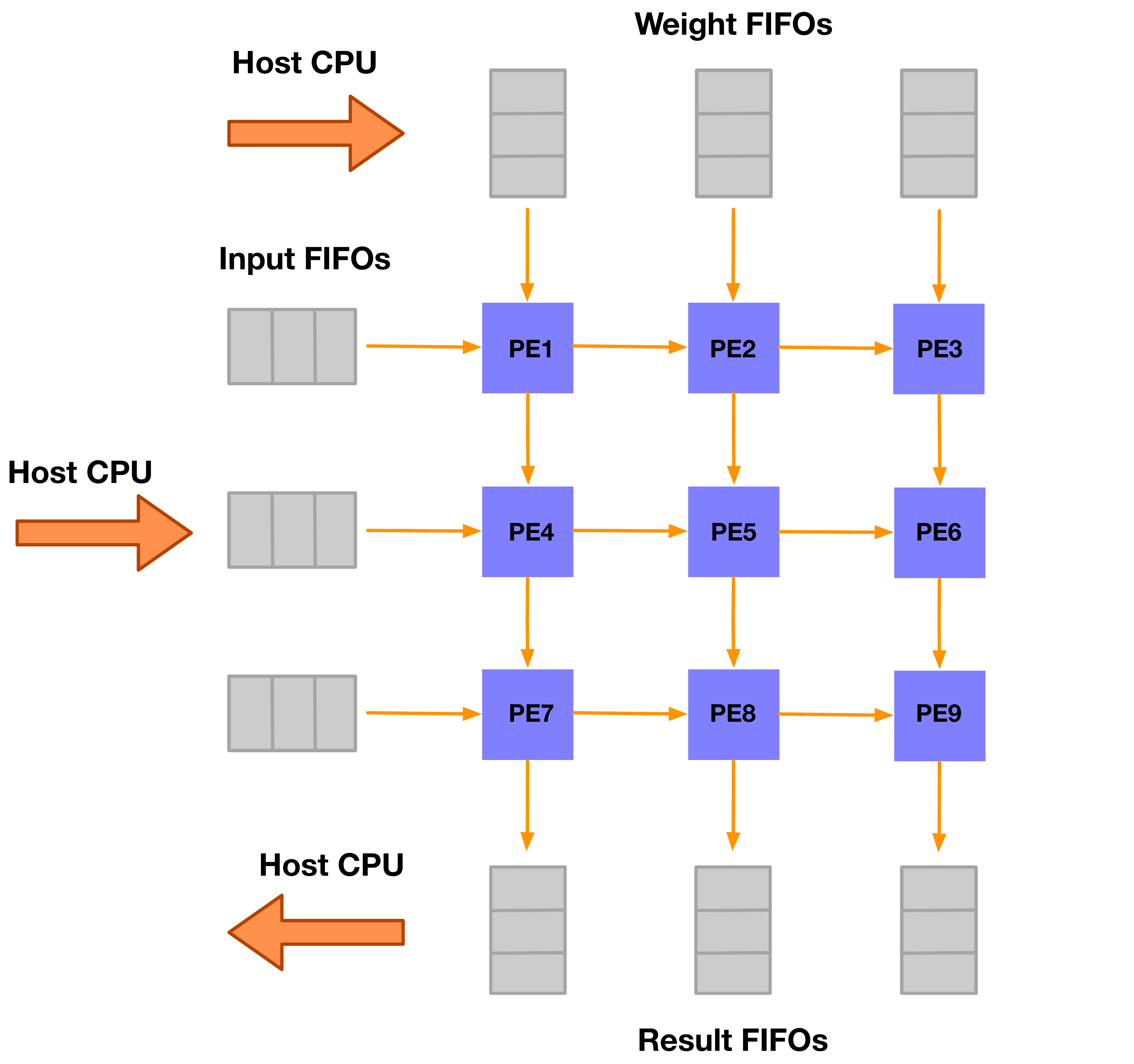}
  \caption{A 2D spatial matrix-vector multiplication array.}
  \label{mv_accel}
\end{figure}

\begin{figure}[t!]
  \includegraphics[width=0.7\linewidth]{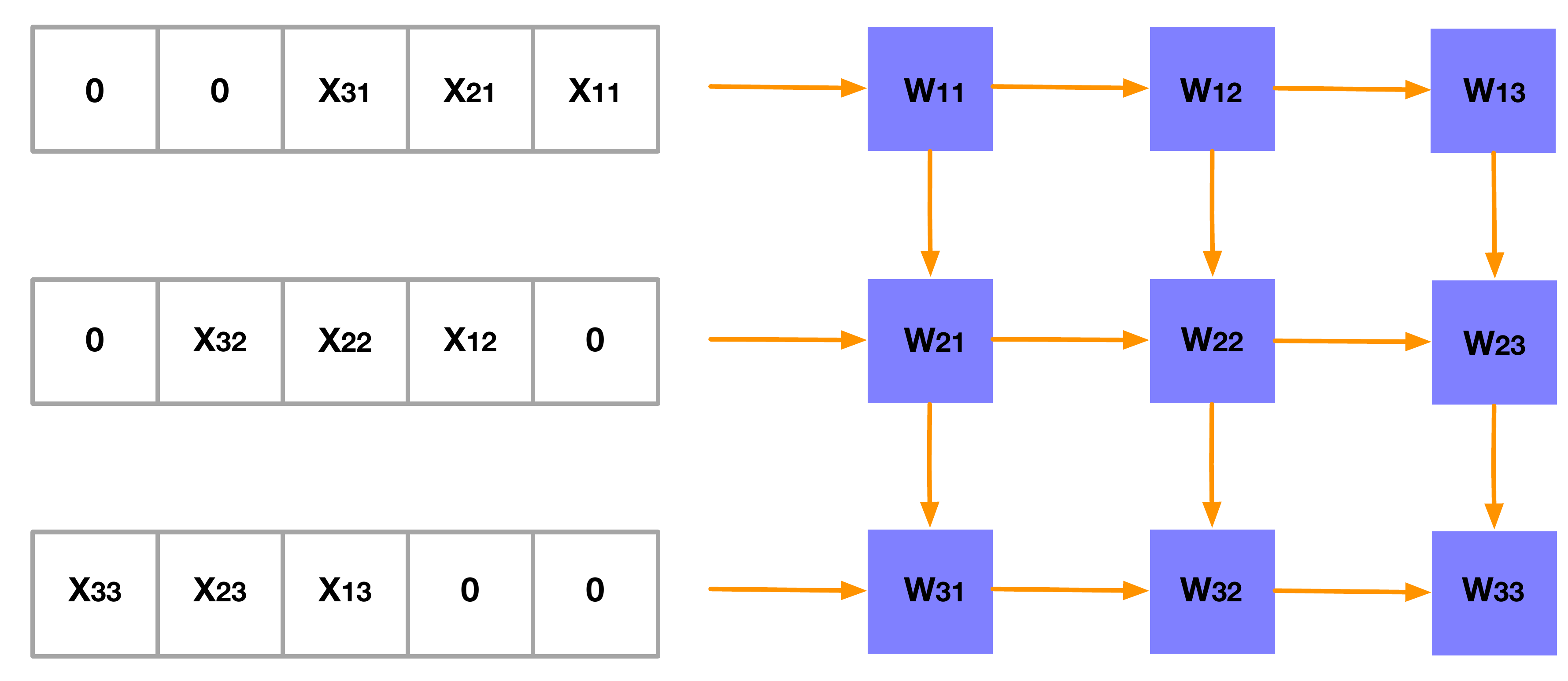}
  \caption{The diagonal wave-front propagation of inputs in the matrix-vector multiplication accelerator.}
  \label{input_mv_accel}
\end{figure}

The matrix-vector multiplication accelerator is a 2D array with multiple PE columns, as shown in Fig. \ref{mv_accel}. The weight and result FIFOs, of the same size as before, are implemented for each PE column. The PE design is unchanged, and propagates inputs from left to right, across PE columns. As before, the accelerator receives inputs and weights from the host CPU and loads the weight matrix into the 3 $\times$ 3 systolic array. During computation, the inputs are arranged in a diagonal wave-front format, Fig. \ref{input_mv_accel}. For each host-to-accelerator data transfer, the matrix-vector multiplication of 3 3 \(\times\) 1 input vectors is computed, taking 7 clock cycles. The host CPU reads out the content of the result FIFOs.

\section{DPA-Style Attack: Setup}

In this section, we demonstrate a DPA on the dot product and MVM accelerators.

\subsection{Threat Model}

We consider two types of attackers. The first type can only observe the inputs to the target device and measure power signatures of the device \cite{dpa_aes_microcontroller}. The second type has a full access to an identical device with the ability to modify the secret values and collect power profiles to facilitate attacks on the target device \cite{template_attacks}, \cite{profile_attack}. We also assume that implementation details of the target device, with the exception of secret weights, are known to the attacker \cite{masked_net}. This is reasonable since the neural network model is often trained after a large effort while the implementation of hardware is usually widely known (public).

\subsection{Experimental Setup}

We synthesized and implemented the dot product and matrix-vector multiplication accelerators, individually, on the SAKURA-G FPGA board. The multiplier and adder of each PE are implemented using a DSP slice. On the FPGA, we floor-planned multiple square physical regions next to one another, assigned one region to each PE, and constrained the implementation of each PE to utilize the resources within its corresponding region. We assigned a trigger signal, which indicates the start of computation, to one of the user GPIO pins of the FPGA board. We used the PicoScope 2408B to capture power traces. FPGA clock frequency is set to 1.5MHz. The power traces are collected at 500MS/s. We choose a low FPGA clock frequency due to sampling rate limit (1GS/s maximum) of the oscilloscope.

\begin{figure}[t!]
  \includegraphics[width=0.8\linewidth]{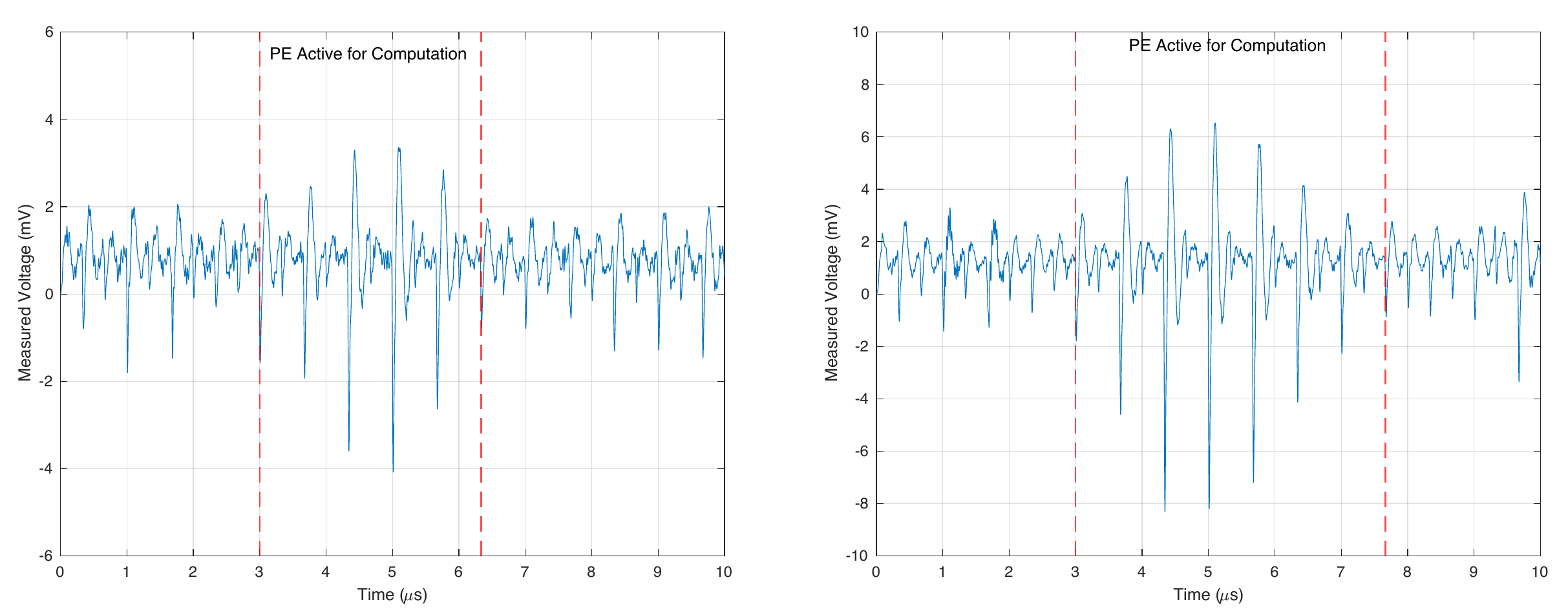}
  \caption{Measured power traces processed by the low pass filter. Left: A trace from the dot product accelerator. Right: A trace from the matrix-vector multiplication accelerator. }
  \label{sample_power_trace}
\end{figure}

Fig. \ref{sample_power_trace} shows one power trace for the dot product and matrix-vector multiplication accelerators. We use a 5-th order low-pass Chebyshev Type I filter with the 0.002 dB passband ripple and a pass-edge frequency of 25MHz to process the raw power traces. The red dash line highlights the clock cycles where PEs are involved in a MAC operation on inputs and weights. These clock cycles are the focus of the DPA. The voltage pulses in the middle of the highlighted region have a larger peak-to-peak value compared to other pulses, since more PEs are active at each MAC operation in these clock cycles due to the diagonal wave-front of inputs. We note that the MVM accelerator shows larger voltage pulses, because a larger systolic array is used.

\section{DPA on Dot Product Accelerator}

We demonstrate how the DPA framework can be modified to successfully attack the dot product accelerator. The strategy allows the attack to succeed in fully identifying the weight vector pre-loaded on the accelerator. We demonstrate the details of the modified DPA framework that utilizes the dependence of the instantaneous MAC output on the earlier-processed weights. Without such a modification, the naive extension of the DPA, as it is used in attacks on common cryptographic ciphers, e.g. AES, fails.

We start by revisiting the algorithm for attacking AES \cite{cpa}:

\begin{itemize}
    \item The attacker collects $N$ power traces with $T$ samples per trace. Each trace corresponds to one encryption on a random input plaintext. The attacker arranges the collected power traces into a $N \times T$ matrix.
    \item For a target AES key byte, and for each AES encryption, the attacker calculates a hypothetical power value based on all 256 possible values of the key byte. The calculated hypothetical power values are arranged in a $256 \times N$ matrix.
    \item The attacker calculates Pearson correlation between each row of the power model matrix and each column of the power trace matrix. The correlation coefficients are arranged in a $256 \times T$ matrix. The row index with the largest value of the correlation matrix represents the attacker's best guess of the target key byte.
\end{itemize}

In the DPA on AES, each key byte of the entire key is extracted individually. The attacker typically focuses on the last round of AES encryption. We propose the following strategy for an attack on a DNN. The attack focuses on a single 8-bit weight at a time. The attacker first collects $N$ power traces, each corresponding to a single dot product computation. Then, the attacker locates the clock cycles in the power traces where the MAC operation involving the weight occurred. We propose that the power model be based on the Hamming distance of the register holding the MAC result between consecutive inputs for each weight. The DPA starts analysis from the first weight in the weight vector and targets weights serially.

First, we follow the above framework directly. We collect 100K power traces from the FPGA board, each corresponding to a dot product computation of three random $3 \times 1$ input vectors (the input FIFO depth is 3) and the fixed (secret) $3 \times 1$ weight vector. We start by focusing on the first weight $w_{11}$, which is associated with the first PE in the column. We use the Hamming distance of \emph{Reg C} of PE1, over the first two consecutive inputs to form the power model. If we denote the two inputs as $x_{11}$ and $x_{21}$, the power model $H_{11}$ is:

\begin{equation}
    H_{11} = HW[(x_{11} \times w_{11}) \oplus (x_{21} \times w_{11})]
    \label{eq1}
\end{equation}

We use the phase of the power traces that corresponds to the switching of the target register and organize data into a $100000 \times 334$ matrix. Then, for each possible value of $w_{11}$, we calculate $H_{11}$. The resulting power models are arranged in a $256 \times 100000$ matrix. Finally, we correlate each row of the power model matrix with each column of the power trace matrix, and select the value of $w_{11}$ with the largest correlation as the attack guess.

\begin{figure*}
\begin{minipage}[b]{0.325\linewidth}
\includegraphics[width=\linewidth]{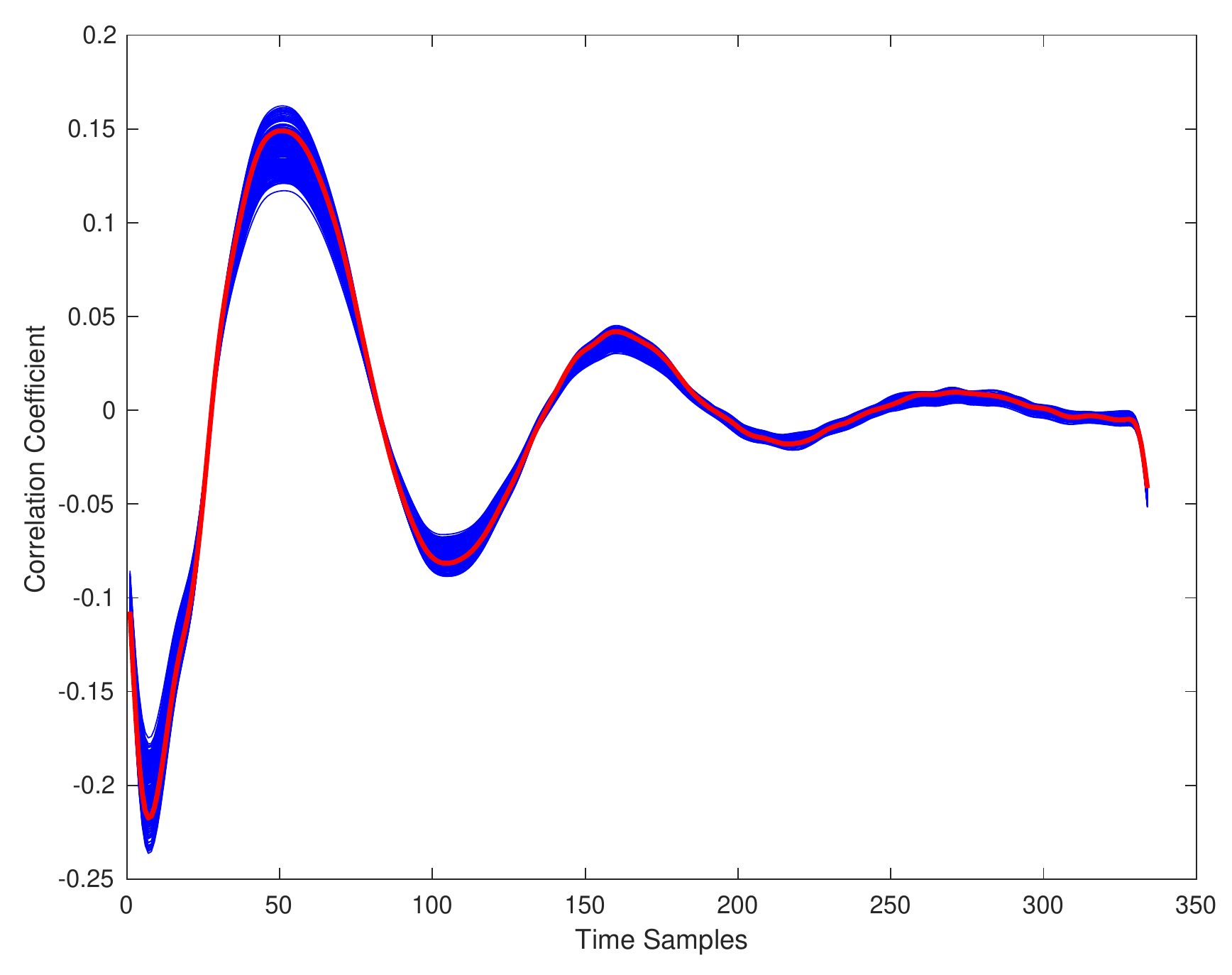}\\
\subcaption{DPA on $w_{11}$.}\label{dpa_dp_w11}
\end{minipage}
\begin{minipage}[b]{0.325\linewidth}
\includegraphics[width=\linewidth]{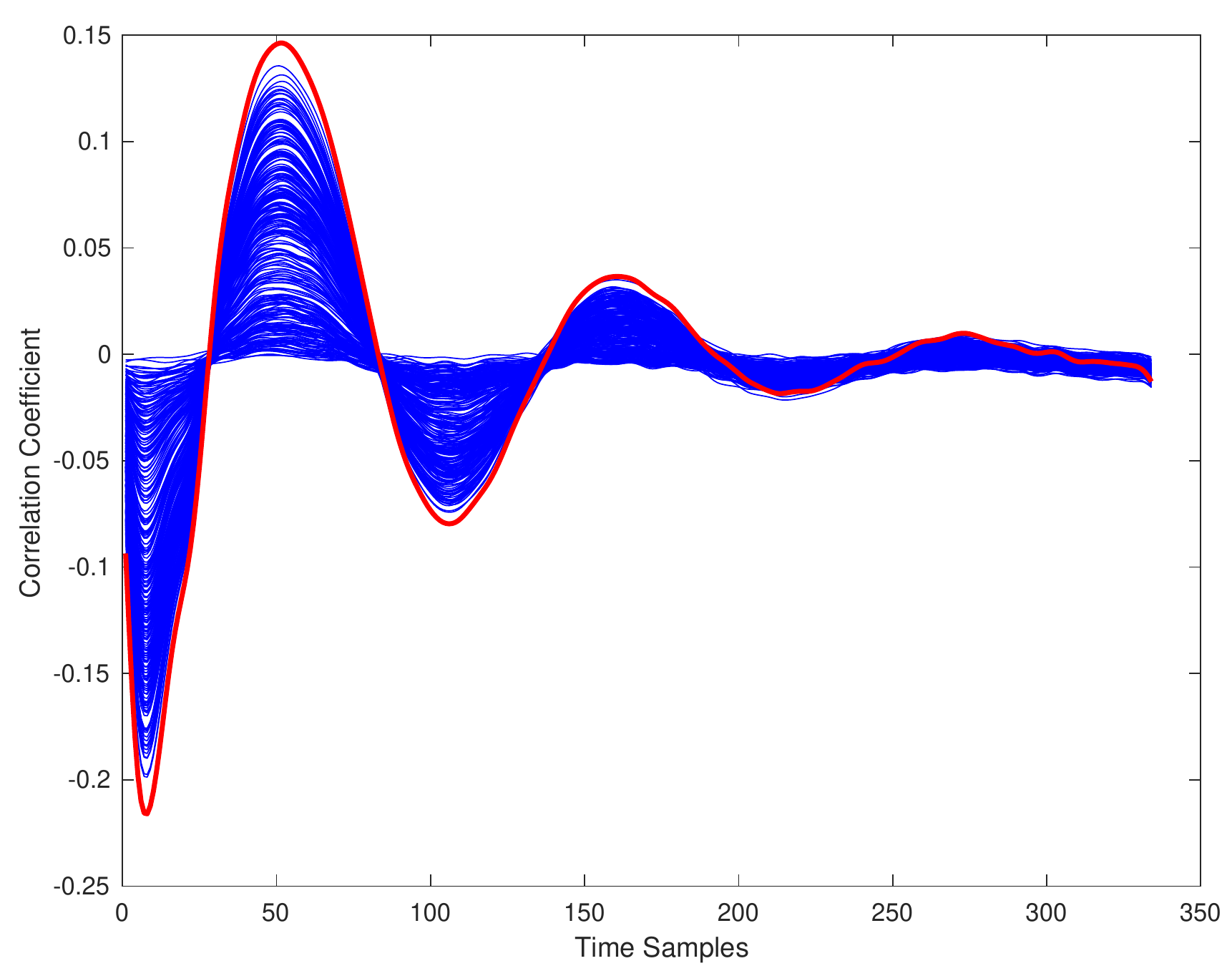}\\
\subcaption{DPA on $w_{21}$.}\label{dpa_dp_w21}
\end{minipage}
\begin{minipage}[b]{0.325\linewidth}
\includegraphics[width=\linewidth]{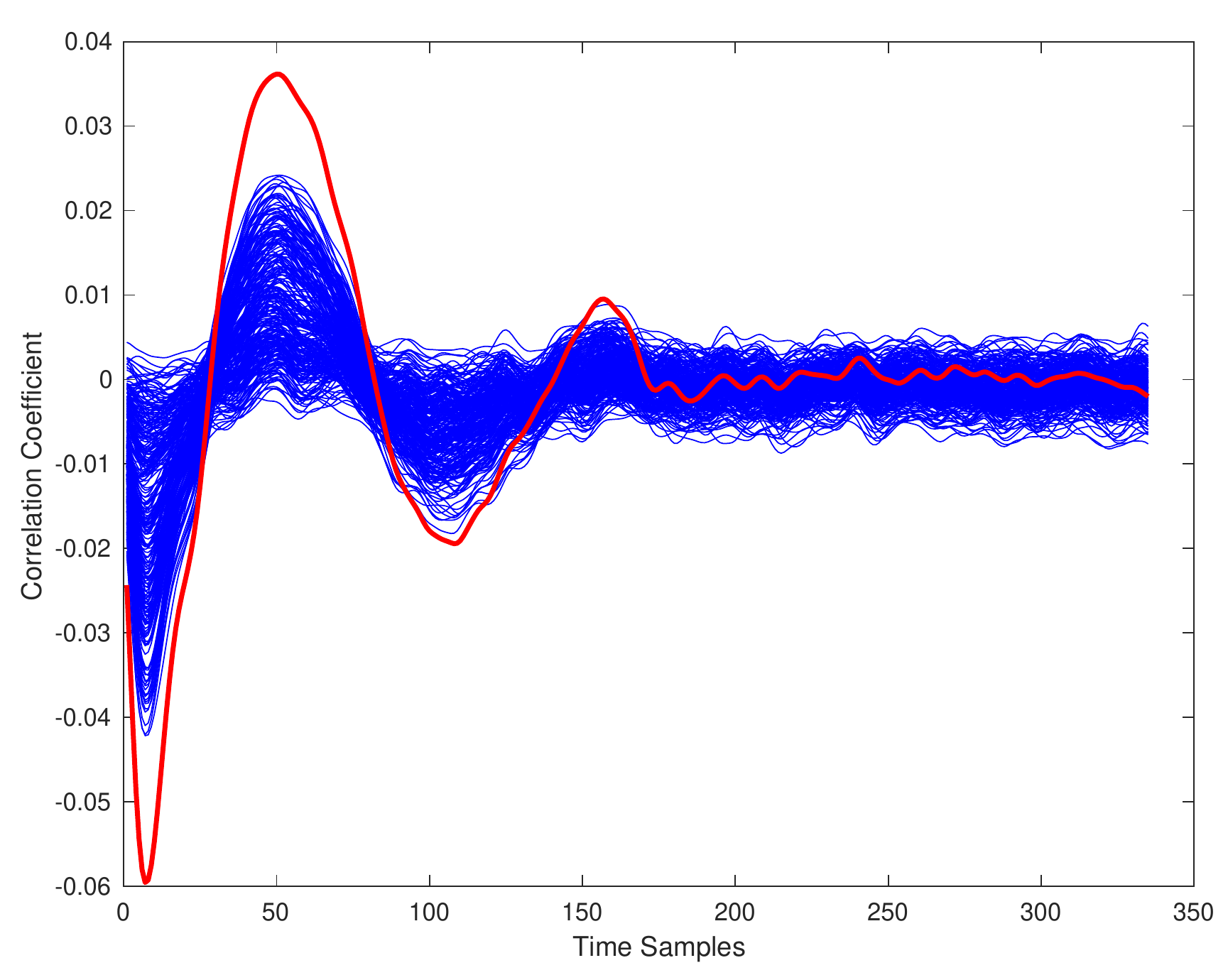}\\
\subcaption{DPA on $w_{31}$.}\label{dpa_dp_w31}
\end{minipage}
\caption{DPA results with 100K power traces. DPA fails to retrieve correct $w_{11}$ value directly, (a). If assume $w_{11}$ is known, DPA succeeded to further retrieve $w_{21}$ and $w_{31}$, (b) and (c).  Red curves represent correct values: $w_{11} = 120$, $w_{21} = 73$ and $w_{31} = -96$.}

\label{dpa_dp_all}
\end{figure*}

Unfortunately, this direct procedure fails to retrieve the correct value of $w_{11}$, Fig. \ref{dpa_dp_w11}. We describe a solution for finding $w_{11}$ later. We first describe the strategy for extracting other weights, assuming $w_{11}$ has already been retrieved and the partial sums produced by $w_{11}$ can be calculated. We now construct the power model $H_{21}$ for $w_{21}$ as: 

\begin{equation}
    H_{21} = HW[({\sum}_{i=1}^2 x_{1i} \times w_{i1}) \oplus ({\sum}_{i=1}^2 x_{2i} \times w_{i1})]
    \label{eq2}
\end{equation}

In Equation \ref{eq2}, $x_{12}$ and $x_{22}$ are the first two inputs within the input FIFO for $w_{21}$. We correlate the power models with the targeted portion of the power traces. The correlation coefficients for all possible values of $w_{21}$  are shown in Fig. \ref{dpa_dp_w21}. This time the guess for $w_{21}$ is based on the highest correlation corresponding to the correct value. The power model $H_{31}$ for the next weight $w_{31}$ is:

\begin{equation}
    H_{31} = HW[({\sum}_{i=1}^3 x_{1i} \times w_{i1}) \oplus ({\sum}_{i=1}^3 x_{2i} \times w_{i1})]
    \label{eq3}
\end{equation}

Here, $x_{13}$ and $x_{23}$ are the first two consecutive input values into the input FIFO for $w_{31}$. The correlation for $w_{31}$ is in Fig. \ref{dpa_dp_w31}. The attack again succeeds to retrieve the correct value of $w_{31}$.

\begin{figure}[t!]
  \includegraphics[width=0.8\linewidth]{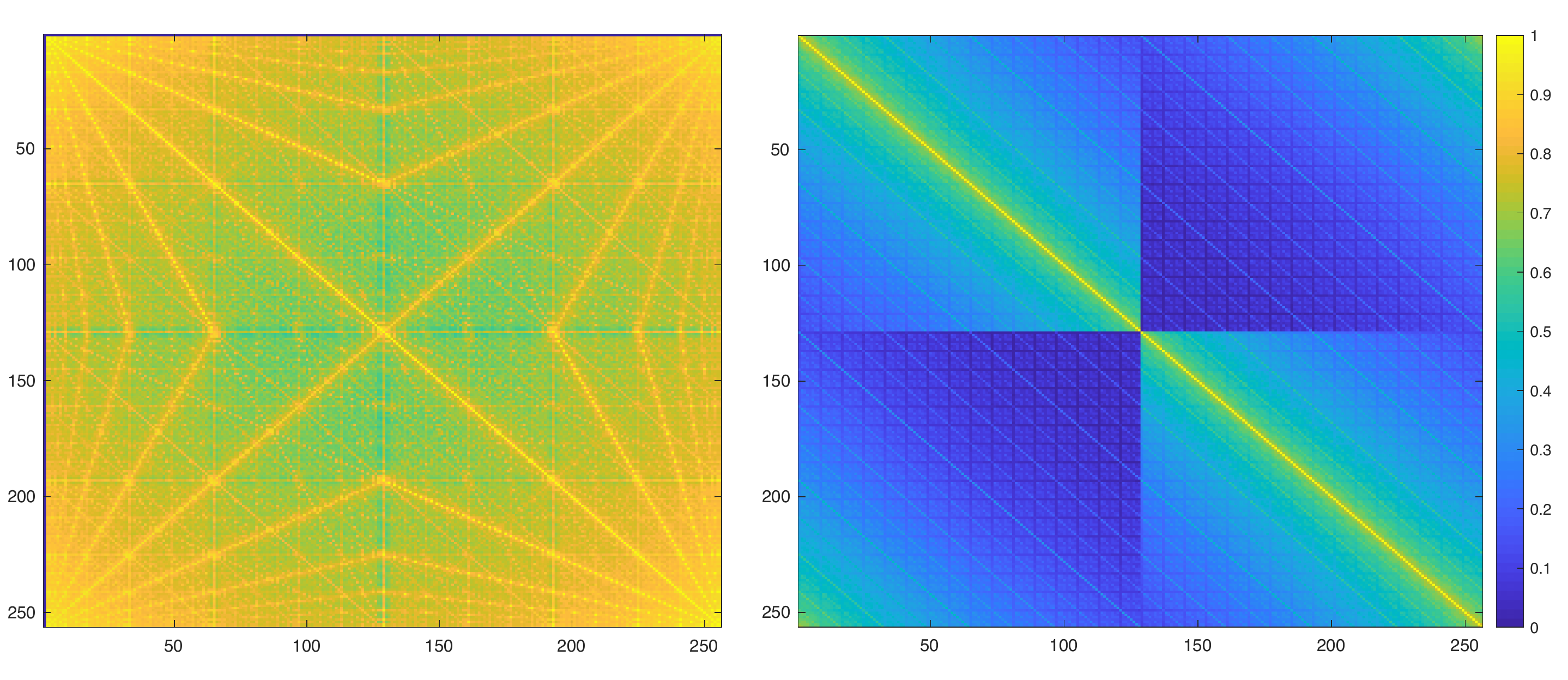}
  \caption{Pairwise correlation across power models of all possible guesses on $w_{11}$ (multiplication only, left) and $w_{21}$ (multiplication and accumulation, right). 
  }
  \label{pair_wise_corr_mult_mac}
\end{figure}

Based on the above discussion the weight $w_{11}$ seems to be the bottleneck: $w_{11}$ needs to be retrieved first in order to retrieve the subsequent weights $w_{21}$ and $w_{31}$. It is critical to understand why the DPA fails to retrieve $w_{11}$ directly. We note that $w_{11}$ is the first element in the weight vector. This means that the previous partial sum input to PE1 is zero. Therefore, the MAC associated with $w_{11}$ involves only multiplication. In contrast, the MACs related to $w_{21}$ and $w_{31}$ involve both multiplication and accumulation. 

DPA examines the correlation between power traces and power models. It is critical that the hypothetical power models based on the incorrect guesses of the secret do not show correlation with the model based on the correct guess. If there is aliasing (correlation between the power model of the correct guess and that of a different guess), it will cause the incorrect guess to also show high correlation. This effectively reduces the confidence in the correct guess. In block ciphers, such as AES, this issue does not arise due to the fundamental non-linearity of the S-box \cite{dpa_and_sbox}. 

In our case, aliasing occurs. We investigate this further to understand the difference in the behavior of $w_{11}$ and $w_{21}$. We calculate pairwise correlation across 256 rows of the power model matrix for $w_{11}$, based on the DPA described above, and repeat the calculation for $w_{21}$. Fig. \ref{pair_wise_corr_mult_mac} shows the results plotted as 2D color maps. It can be seen that power models of different guesses on $w_{11}$ show large correlation indicating large aliasing. Power models of different guesses of $w_{21}$ show only a large self-correlation. Note that the difference is due to the absence of accumulation. (In the case of $w_{11}$ extraction, the role of accumulation is intriguing and we plan to explore it further in the future.)

To overcome the difficulty of retrieving $w_{11}$ directly, we use the MACs involving additional, subsequent weights to extract the correct $w_{11}$. 
Since the product generated by $w_{11}$ also determines the MAC outputs for $w_{21}$ and $w_{31}$, the hypothetical power model will show high correlation only for the correct weight combinations. To achieve this, we modify the attack procedure to be:

\begin{itemize}
    \item Perform DPA on $w_{11}$. Construct the power model matrix using Equation \ref{eq1} and the power trace matrix. Calculate the correlation between each row of the power model matrix and each column of the power trace matrix. Sort all guesses on $w_{11}$ based on the  maximum correlation found in the time window. Since the rank of correct guess on $w_{11}$ is close to 50, we record the $w_{11}$ guesses corresponding to 50 highest correlations. 
    \item For each recorded guess on $w_{11}$, perform DPA on $w_{21}$. Construct the power model matrix using Equation \ref{eq2} and the power trace matrix and calculate the correlations. Record the guess on $w_{21}$ with the highest correlation and its corresponding $w_{11}$ guess. This results in 50 $(w_{11}, w_{21})$ guesses.
    \item With each recorded $(w_{11}, w_{21})$ guess, perform DPA on $w_{31}$. Construct the power model matrix using Equation \ref{eq3} and the power trace matrix and calculate the correlations. Record the guess on $w_{31}$ with the highest correlation and its corresponding $(w_{11}, w_{21})$ guess. Return the $(w_{11}, w_{21}, w_{31})$ guess with the highest correlation among all 50 recorded combinations as the final guess.  
\end{itemize}

\begin{figure}
\begin{minipage}[b]{0.4\linewidth}
\includegraphics[width=\linewidth]{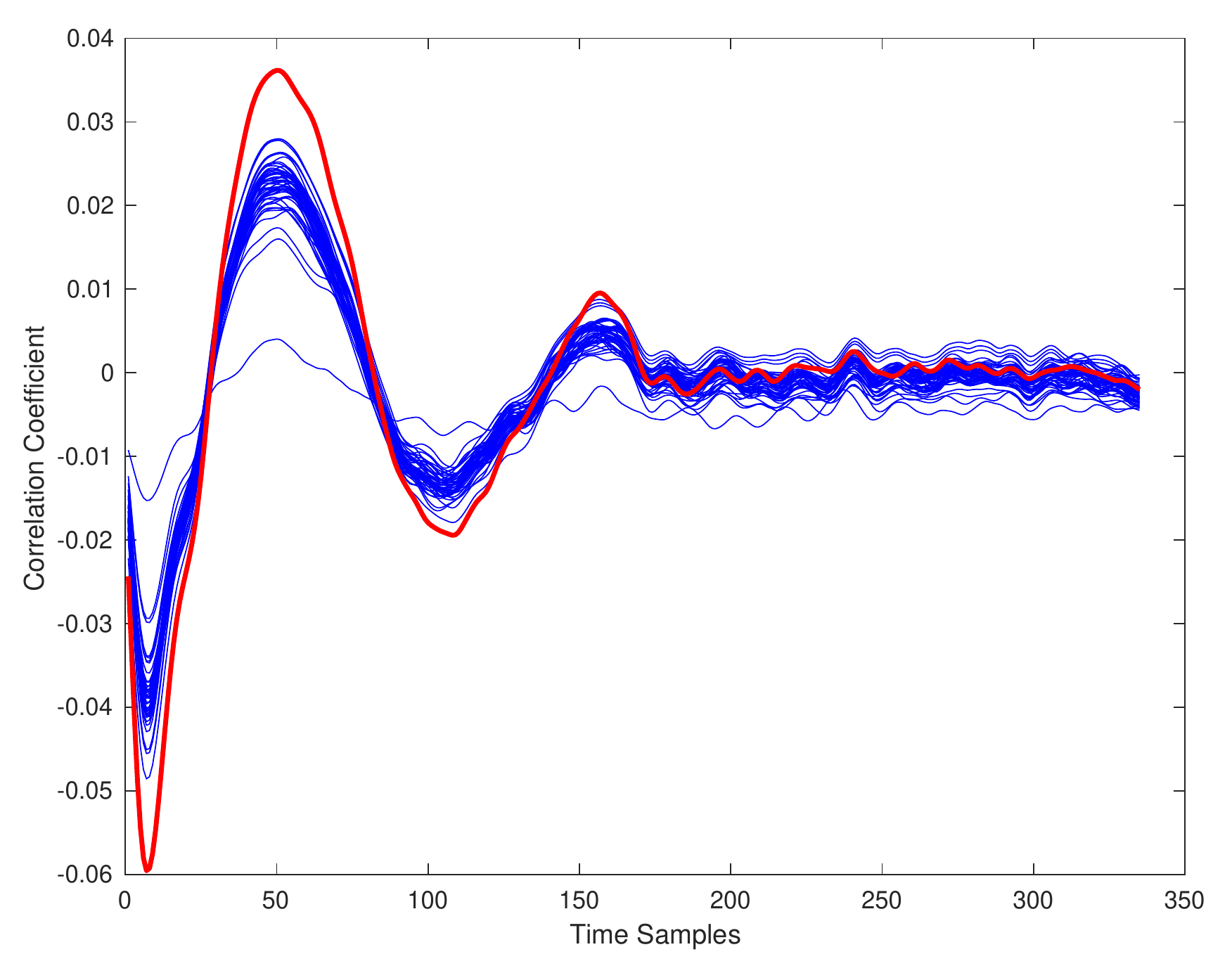}\\
\subcaption{DPA with 100K traces.}\label{dp_compound_dpa_100k}
\end{minipage}
\begin{minipage}[b]{0.4\linewidth}
\includegraphics[width=\linewidth]{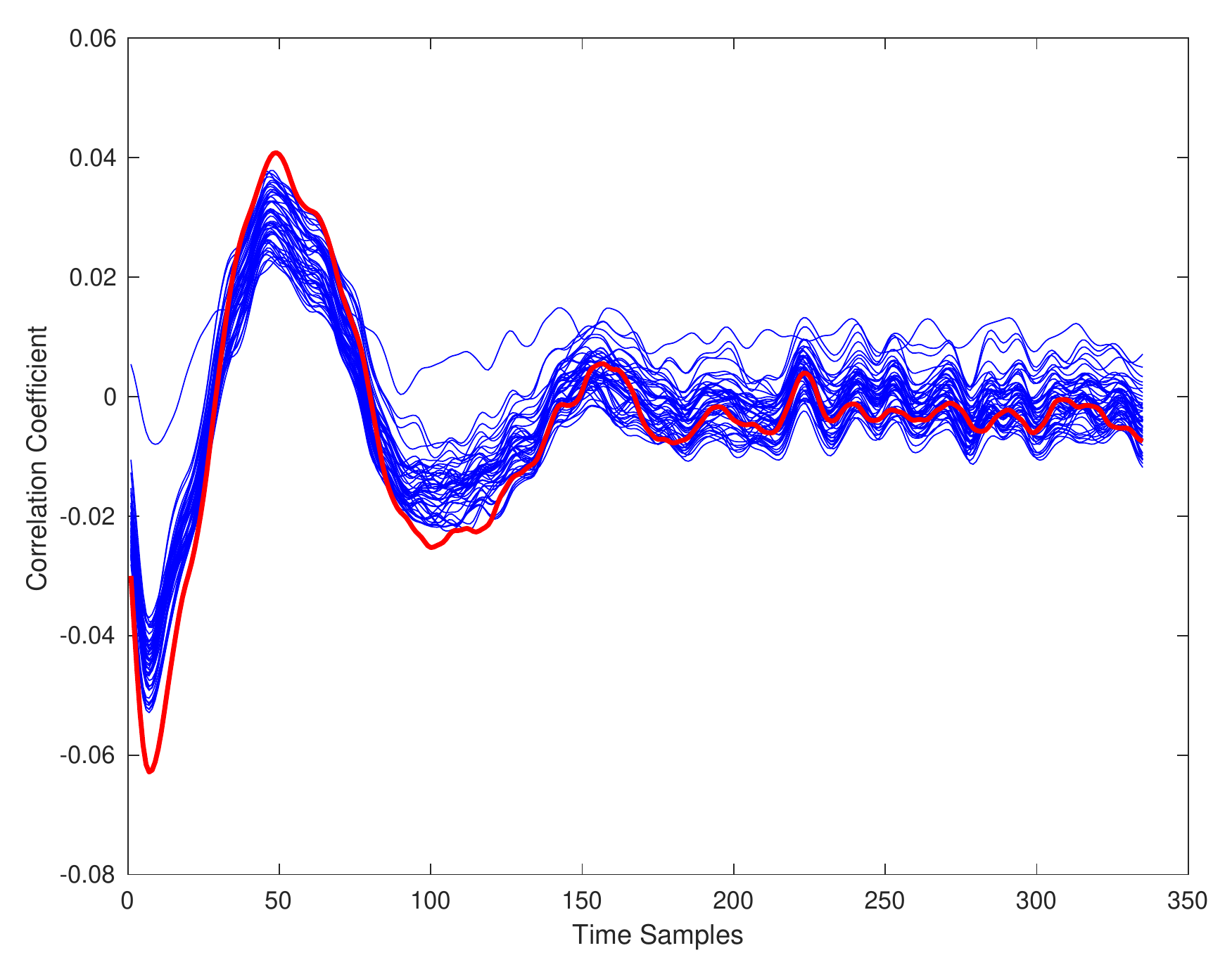}\\
\subcaption{DPA with 20K traces.}\label{dp_compound_dpa_20k}
\end{minipage}

\caption{Modified DPA framework successfully retrieves $(w_{11}, w_{21}, w_{31})$ with 100K power traces, (a). 20K traces is identified as the minimum number of traces required for a successful attack, (b). Red curves represent the correct combination: $(w_{11}, w_{21}, w_{31}) = (120, 73, -96)$.}

\label{dp_compound_dpa}
\end{figure}

The time complexity of the modified DPA framework depends on how many guesses of $w_{11}$ are recorded. The number of $w_{11}$ guesses to record leads to a linear increase in computation complexity. We apply the above modified DPA framework to 100K power traces. The correct guess on $(w_{11}, w_{21}, w_{31})$ is successfully retrieved, as shown in Fig. \ref{dp_compound_dpa_100k}. We start with 10K power traces and increase the number of traces used in steps of 10K, repeating the experiments. We identify 20K traces as the Measurement to Disclosure (MTD) for the dot product accelerator. The correlation plot of DPA with 20K power traces is shown in Fig. \ref{dp_compound_dpa_20k}. We summarize the results of the attack on the dot product accelerator in Table \ref{tab:dpa_dp_results}. The Rank column shows the rank of the correct guess on $(w_{11}, w_{21}, w_{31})$ among all 50 recorded combinations and the Correlation column shows the Pearson correlation of the correct guess at 20K (MTD) power traces.

\begin{table}[htbp]
    \centering
    \begin{tabular}{|c|c|c|c|c|}
    \hline
        Weight & Correct Guess & Rank & Correlation & MTD  \\
    \hline
        $(w_{11}, w_{21}, w_{31})$ & (120, 73, -96) & 1 & 0.0628 & 20000 \\ 
    \hline
    \end{tabular}
    \caption{DPA results on the dot product accelerator. The correct weights can be retrieved with 20K traces.}
    \label{tab:dpa_dp_results}
\end{table}

\section{Higher Security of Matrix-Vector Multiplication Accelerator}

In this section, we study the vulnerability of a 2D matrix-vector multiplication accelerator to a DPA-style attack. We find that the 2D accelerator exhibits higher security. Both a conventional DPA and a stronger template-based DPA fail. We investigate the causes of higher resistance of the 2D array to the attack compared to the dot product accelerator. \emph{We explain the reason for higher security of a 2D array design by the fact that it intrinsically results in multiple instantaneous MAC outputs being dependent on the same input.} We show that this is a fundamental feature of the commonly-used weight-stationary dataflow. 

We first investigate the conventional DPA attack that uses the approach described above. A conventional DPA does not require access to an identical profiled device. We collect 460K power traces from the matrix-vector multiplication accelerator. (We stopped at 460K traces due to measurement time budget.) For each PE column, the top 50 guesses of the first weight of each column ($w_{11}$ for column 1, $w_{12}$ for column 2, $w_{13}$ for column 3) are recorded and the relevant phases of power traces are selected. However, the conventional DPA fails to retrieve the weights, as shown in Table \ref{tab:dpa_mv_results_multi}, where NA indicates that the correct weight combination does not appear in the 50 recorded weight pairs.

\begin{table}[htbp]
    \centering
    \begin{tabular}{|c|c|c|c|c|}
    \hline
         Weight & Correct Guess & Rank & Correlation & MTD  \\
    \hline
         $(w_{11}, w_{21}, w_{31})$ & (23, -107, 74) & NA & NA & NA \\
    \hline
         $(w_{12}, w_{22}, w_{32})$ & (120, 73, -96) & NA & NA & NA \\
    \hline 
         $(w_{13}, w_{23}, w_{33})$ & (-6, -31, 17) & 1 & 0.0640 & 20000 \\ 
    \hline
    \end{tabular}
    \caption{Results of conventional DPA on the matrix-vector multiplication accelerator. Conventional DPA fails to retrieve $(w_{11}, w_{21}, w_{31})$ and $(w_{12}, w_{22}, w_{32})$ with 460K power traces: the correct weight combination does not even appear in the 50 recorded weight pairs. $(w_{13}, w_{23}, w_{33})$ can be retrieved with 20K traces.}
    \label{tab:dpa_mv_results_multi}
\end{table}

\emph{We now investigate a template-based DPA, which assumes a stronger adversary.} Since DPA relies on the analysis of power consumed by MACs, to improve the effective SNR, we propose a profiling technique that removes all non-MAC power. We assume an attacker has full access to an identical device, which can be used for profiling. Hence, the attacker can modify the secret weights of the profiled device and collect its power traces.  Specifically, the attacker sets all the weights of the profiled device to zero. A trace (template) from the profiled device captures the systolic array power minus the MAC power. The attacker produces a power template for each observed input. This means the same number of template power traces need to be collected from the profiled device as the target power traces. The attacker subtracts the template power trace from the target power trace, which is then used for DPA. We note that the proposed attack is different from the template attack of Chari et al. \cite{template_attacks}, which constructs a template using the mean trace and the noise covariance matrix for each key value. We call our attack the template-based DPA to reflect the fact that an identical device is used for profiling.

\begin{table}[htbp]
    \centering
    \begin{tabular}{|c|c|c|c|c|}
    \hline
         Weight & Correct Guess & Rank & Correlation & MTD  \\
    \hline
         $(w_{11}, w_{21}, w_{31})$ & (23, -107, 74) & 32 & 0.0623 & NA \\
    \hline
         $(w_{12}, w_{22}, w_{32})$ & (120, 73, -96) & NA & NA & NA \\
    \hline 
         $(w_{13}, w_{23}, w_{33})$ & (-6, -31, 17) & 1 & 0.0633 & 20000 \\ 
    \hline
    \end{tabular}
    \caption{Results of the template-based DPA on matrix-vector multiplication accelerator. Template-based DPA fails to retrieve $(w_{11}, w_{21}, w_{31})$ and $(w_{12}, w_{22}, w_{32})$ with 460K traces. The correct combination for $(w_{11}, w_{21}, w_{31})$ appears in the 50 recorded weight pairs but does not show the highest correlation. $(w_{13}, w_{23}, w_{33})$ can be retrieved with 20K traces.}
    \label{tab:dpa_mv_results_multi_profile}
\end{table}

The attack just described allows extracting some, but not all, weights, even after collecting a much larger number of traces (460K traces). Table \ref{tab:dpa_mv_results_multi_profile} shows that the template-based DPA fails to retrieve the weights of 2 out of 3 columns. 

We verify our conclusions on the 2D matrix-vector multiplication accelerator by performing the attack using simulated noise-free power traces. We generate the simulated traces by modeling the register switching in each PE. The power consumption of each PE at a specific clock cycle is modeled as:

\begin{equation}
    P_x = \alpha \cdot HD(Reg\ A) + \beta \cdot HD(Reg\ C)
    \label{eq4}
\end{equation}

The power contribution of \emph{Reg\ A} and \emph{Reg\ C} defines each term of the equation. $HD(\cdot)$ denotes the Hamming distance of the register values over two consecutive clock cycles. Coefficients $\alpha$ and $\beta$ are used to adjust the contributions of registers in different PEs, based on their load capacitance. Specifically, for PE1, PE2, PE4, PE5, PE7 and PE8, we use $\alpha = 3$ and $\beta = 2$; for PE3, PE6 and PE9, we use $\alpha = 1$ and $\beta = 2$. We sum the power of each PE to get the total power at a specific clock cycle. This represents the power averaged over one clock cycle. We repeat the register-switching calculation for each compute-active clock cycle of the 2D systolic array, obtaining a simulated trace with 7 samples.

We generate 460K simulated noise-free traces using the same inputs as the 460K measured traces. We perform both the conventional DPA and template-based DPA with the simulated traces and summarize the results in Table \ref{tab:dpa_mv_results_multi_sim} and \ref{tab:dpa_mv_results_multi_profile_sim}.

\begin{table}[htbp]
    \centering
    \begin{tabular}{|c|c|c|c|c|}
    \hline
         Weight & Correct Guess & Rank & Correlation & MTD  \\
    \hline
         $(w_{11}, w_{21}, w_{31})$ & (23, -107, 74) & 45 & 0.2627 & NA \\
    \hline
         $(w_{12}, w_{22}, w_{32})$ & (120, 73, -96) & NA & NA & NA \\
    \hline 
         $(w_{13}, w_{23}, w_{33})$ & (-6, -31, 17) & 1 & 0.3878 & 10000 \\ 
    \hline
    \end{tabular}
    \caption{Results of conventional DPA on matrix-vector multiplication accelerator with simulated noise-free traces. Only $(w_{13}, w_{23}, w_{33})$ are retrieved, which matches results in Table \ref{tab:dpa_mv_results_multi}.}
    \label{tab:dpa_mv_results_multi_sim}
\end{table}

\begin{table}[htbp]
    \centering
    \begin{tabular}{|c|c|c|c|c|}
    \hline
         Weight & Correct Guess & Rank & Correlation & MTD  \\
    \hline
         $(w_{11}, w_{21}, w_{31})$ & (23, -107, 74) & 35 & 0.2952 & NA \\
    \hline
         $(w_{12}, w_{22}, w_{32})$ & (120, 73, -96) & NA & NA & NA \\
    \hline 
         $(w_{13}, w_{23}, w_{33})$ & (-6, -31, 17) & 1 & 0.4420 & 10000 \\ 
    \hline
    \end{tabular}
    \caption{Results of template-based DPA on matrix-vector multiplication accelerator with simulated noise-free traces. Contribution of \emph{Reg A} is removed from the simulated traces. Only $(w_{13}, w_{23}, w_{33})$ are retrieved, which matches results in Table \ref{tab:dpa_mv_results_multi_profile}.}
    \label{tab:dpa_mv_results_multi_profile_sim}
\end{table}

We also explored the Hamming weight power models and bit-level power models in the template-based DPA. For a specific PE at a specific clock cycle, the Hamming weight power model is constructed as the Hamming weight of \emph{Reg C}. The bit-level power model is constructed as the Hamming distance of a single bit of \emph{Reg C} \cite{Mangard_power_analysis_attack} over two consecutive cycles. (We selected one bit to explore the behavior but we believe the results do not depend on which bit is used.) We substitute the power models given by Equation \ref{eq1} to \ref{eq3} with the Hamming weight and bit-level power models and repeat the DPA described above. Unfortunately, \emph{none} of the weights could be retrieved with the new power models even with 460K power traces, as shown in Table \ref{tab:dpa_mv_results_multi_profile_hw_460k} and \ref{tab:dpa_mv_results_multi_profile_bit_460k}. 
We believe that the failure of the Hamming weight power model is due to the inaccurate capture of the PE power consumption while the failure of the bit-level model is due to the interference of switching of different register bits. 

\begin{table}[htbp]
    \centering
    \begin{tabular}{|c|c|c|c|c|}
    \hline
         Weight & Correct Guess & Rank & Correlation & MTD  \\
    \hline
         $(w_{11}, w_{21}, w_{31})$ & (23, -107, 74) & NA & NA & NA \\
    \hline
         $(w_{12}, w_{22}, w_{32})$ & (120, 73, -96) & NA & NA & NA \\
    \hline 
         $(w_{13}, w_{23}, w_{33})$ & (-6, -31, 17) & NA & NA & NA \\ 
    \hline
    \end{tabular}
    \caption{Results of template-based DPA on matrix-vector multiplication accelerator with the Hamming weight power model. None of the weights can be retrieved with 460K traces.}
    \label{tab:dpa_mv_results_multi_profile_hw_460k}
\end{table}

\begin{table}[htbp]
    \centering
    \begin{tabular}{|c|c|c|c|c|}
    \hline
         Weight & Correct Guess & Rank & Correlation & MTD  \\
    \hline
         $(w_{11}, w_{21}, w_{31})$ & (23, -107, 74) & 17 & 0.0080 & NA \\
    \hline
         $(w_{12}, w_{22}, w_{32})$ & (120, 73, -96) & 6 & 0.0100 & NA \\
    \hline 
         $(w_{13}, w_{23}, w_{33})$ & (-6, -31, 17) & 6 & 0.0085 & NA \\ 
    \hline
    \end{tabular}
    \caption{Results of template-based DPA on matrix-vector multiplication accelerator with the bit-level power model. None of the weights can be retrieved with 460K traces.}
    \label{tab:dpa_mv_results_multi_profile_bit_460k}
\end{table}

\emph{The 2D matrix-vector multiplication accelerator appears significantly less vulnerable to a DPA-style attack compared to a simpler 1D array.}
To understand the source of the improved resistance to DPA we conduct additional experiments. We investigate the exploratory case where only a single column of PE of the $3 \times 3$ systolic array is activated. In this case, the accelerator is essentially performing a dot product of the input vector with the weight column. The difference is that inputs still propagate horizontally across the PEs and contribute to power. We implement this case by pre-loading the weights of the target PE column only and pre-loading zero weights to the remaining PE columns. MACs of these columns do not contribute power since all their partial sums remain zero. For consistency with the previous template-based experiment, we assume the attacker has access to an identical device to collect power traces. We perform exploratory study for each individual PE column: the PE column is active while other columns are inactive. In each study, the templates (to be subtracted) are based on 100K traces. Table \ref{tab:dpa_mv_results_single_profile} summarizes the results of the exploratory studies for the individual PE columns. 

\begin{table}[htbp]
    \centering
    \begin{tabular}{|c|c|c|c|c|}
    \hline
         Weight & Correct Guess & Rank & Correlation & MTD  \\
    \hline
         $(w_{11}, w_{21}, w_{31})$ & (23, -107, 74) & 1 & 0.0448 & 30000 \\ 
    \hline
         $(w_{12}, w_{22}, w_{32})$ & (120, 73, -96) & 1 & 0.0359 & 20000 \\ 
    \hline 
         $(w_{13}, w_{23}, w_{33})$ & (-6, -31, 17) & 1 & 0.0651 & 30000 \\ 
    \hline
    \end{tabular}
    \caption{Results of template-based DPA for exploratory studies. Template-based DPA successfully retrieves weights of each individual PE column, while other PE columns are inactive. Corresponding MTDs are identified.}
    \label{tab:dpa_mv_results_single_profile}
\end{table}

The experiments demonstrate that all weights in each PE column were retrieved, needing at most 30K traces. This confirms the vulnerability of the dot product to DPA. \emph{It also proves that simultaneous MACs of different PE columns create a higher resistance to DPA by contributing power interfering with the selection of the correct hypothesis.} 

\begin{figure}[t!]
  \includegraphics[width=0.75\linewidth]{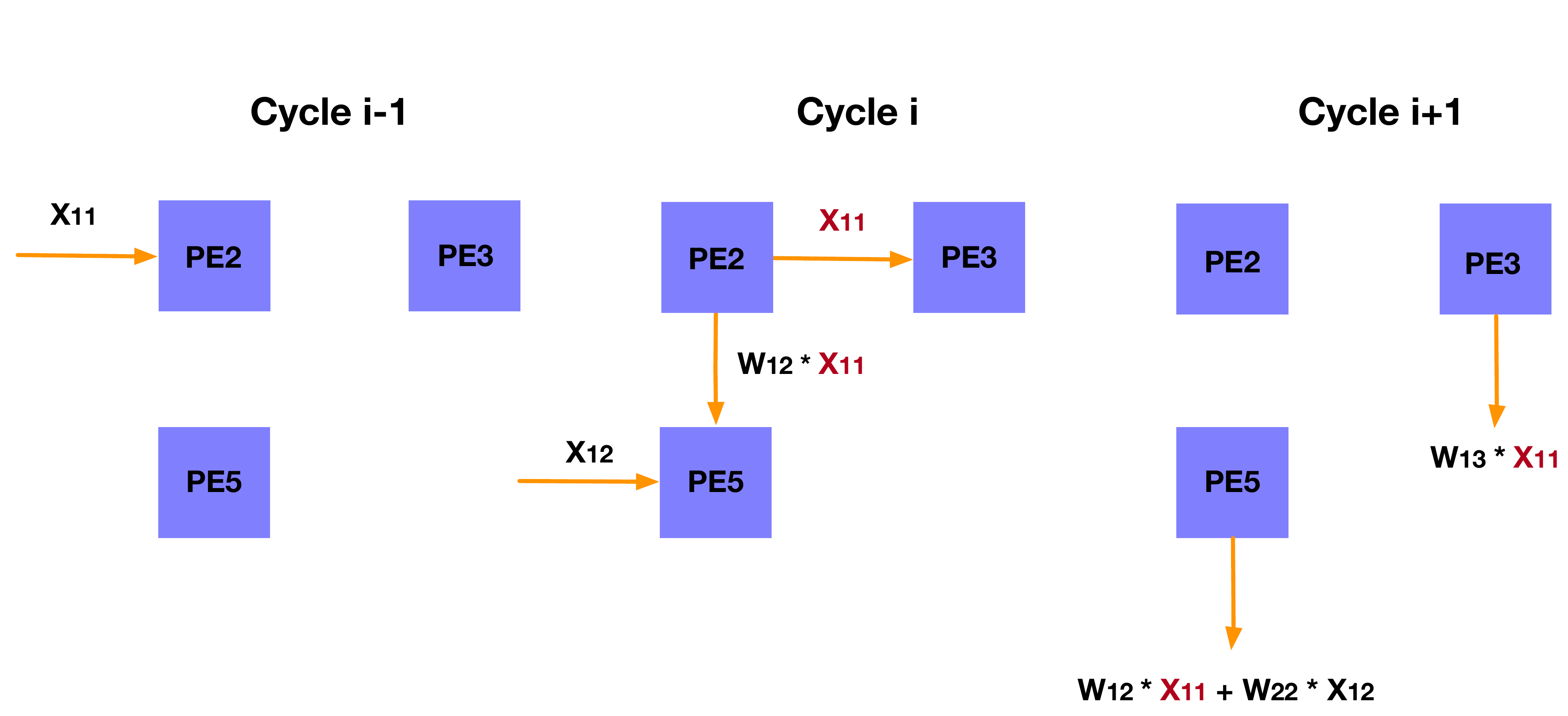}
  \caption{Illustration of multiple simultaneous MAC outputs dependent on the same input with PE2, PE3, and PE5: At cycle $i-1$, input $X_{11}$ propagates to PE2 (Left); At cycle $i$, the MAC output to be computed in PE3 and the previous partial sum to PE5 are determined by the same input $X_{11}$ (Middle); At cycle $i+1$, the MAC output of both PE3 and PE5 depends on input $X_{11}$ (Right).}
  \label{same_input_multi_mac}
\end{figure}

We believe that this behavior, in which MAC operations of different columns cause issues for DPA, is a general characteristic of 2D accelerators based on the weight-stationary dataflow. The behavior is caused by \emph{multiple MAC outputs depending on the same inputs simultaneously.} This is because the weight-stationary dataflow results in inputs to the matrix-vector multiplication array to be arranged in a diagonal wave-front format, as shown in Fig. \ref{input_mv_accel}. The MAC outputs of a PE and the PE on its lower left in its adjacent column (if applicable), will be affected by the same input(s) simultaneously. To illustrate this dataflow feature, we focus on PE2, PE3 and PE5. The MAC output to be computed in PE3, and the previous partial sum to PE5, are determined by the same input at a specific clock cycle. PE5 accumulates its input-weight products with the previous partial sum. This means the same input affects the MAC outputs of both PE3 and PE5 \emph{simultaneously}, Fig. \ref{same_input_multi_mac}. 

We assess the interference by examining the correlation between switching of \emph{Reg C} of different PEs in the same clock cycle. Since \emph{Reg C} holds a MAC output of each PE, the power due to the switching of \emph{Reg C} represents each PE's MAC power for the purpose of DPA correlation analysis. We calculate the pairwise correlation of Hamming distances of \emph{Reg C} in PE1 to PE8, Table \ref{tab:pairwise_corr_regc}. 

\begin{table}[htbp]
    \centering
    \begin{tabular}{|c|c|c|c|c|c|c|c|c|}
    \hline
         & PE1 & PE2 & PE3 & PE4 & PE5 & PE6 & PE7 & PE8 \\
    \hline
        PE1 & 1.00 & 0 & 0 & 0 & 0 & 0 & 0 & -0.01 \\
    \hline
        PE2 & 0 & 1.00 & 0 & 0.01 & 0 & 0 & 0 & 0 \\
    \hline 
        PE3 & 0 & 0 & 1.00 & 0 & 0.29 & 0 & 0.01 & 0 \\
    \hline 
        PE4 & 0 & 0.01 & 0 & 1.00 & 0 & 0 & 0 & 0 \\
    \hline 
        PE5 & 0 & 0 & 0.29 & 0 & 1.00 & 0 & 0.01 & 0 \\
    \hline 
        PE6 & 0 & 0 & 0 & 0 & 0 & 1.00 & 0 & -0.13 \\
    \hline 
        PE7 & 0 & 0 & 0.01 & 0 & 0.01 & 0 & 1.00 & 0 \\
    \hline 
        PE8 & -0.01 & 0 & 0 & 0 & 0 & -0.13 & 0 & 1.00 \\
    \hline
    \end{tabular}
    \caption{Pairwise correlation of Hamming Distance of \emph{Reg C} in PE1 to PE8.} 
    \label{tab:pairwise_corr_regc}
\end{table}

We observe that some PEs exhibit large correlation in addition to self-correlation. Specifically, PEs whose MAC results are simultaneously affected by the same inputs show non-zero correlation: (PE2, PE4), (PE3, PE5), (PE5, PE7) and (PE6, PE8). Some of them, (PE3, PE5) and (PE6, PE8), show high correlation. 
As discussed earlier, such correlation will interfere with DPA's effectiveness and reduce the confidence of the correct hypothesis for a target PE.

In contrast, for a 1D array, such interference does not occur since the inputs stop propagation after being consumed by the PEs and the same input is never used across multiple PEs. The MAC outputs of different PEs in the 1D array, at any clock cycle, are determined by different inputs. Thus, MAC power of different PEs will \emph{not} show correlation as shown by zero correlations between (PE1, PE4, PE7) and between (PE2, PE5, PE8) in Table \ref{tab:pairwise_corr_regc}.

\section{Breaking 2D Accelerator with Enhanced Template DPA }

In this section, we demonstrate a stronger template-based DPA that succeeds in fully retrieving the weights of the 2D array. The attack requires multiple profiling phases. We call it multi-phase template-based DPA.

As discussed, 2D array exhibits higher resistance to DPA due to parallel MAC operations of different PE columns. To fully retrieve the weights, it is critical to focus on each PE column individually and remove the interference of other columns. However, localizing leakage from each PE column is challenging because the power trace captures aggregate power of the entire 2D array. However, it is possible to expose the leakage of each PE column via a sequential analysis. Specifically, we identify the most vulnerable PE column, extract its weights, remove the effect of the column by template, and move to the next most vulnerable remaining PE column. The process can be repeated until the weights from all PE columns are retrieved. The main question is how to find the most vulnerable PE column of the 2D array.

Using previous results, we observe that the rightmost PE column (PE3, PE6 and PE9) appears to be the easiest to break. The attack can break the column with 20K power traces while the other columns remain secure even after 460K power traces. The reason for this behavior is that input propagation stops at the rightmost PE column. We believe that \emph{the PE column furthest from inputs is the most vulnerable one and its weights can be retrieved most easily}. Based on the hypothesis, we propose the following attack. For simplicity, we use the term ``trace'' to refer to a power trace and ``template trace'' to refer to a power trace collected from a profiled device with the fixed weights.

\begin{itemize}
    \item Perform DPA on column 3 (PE3, PE6, PE9). Run DPA for 1D array. Retrieve weights $(w_{13}, w_{23}, w_{33})$.
    \item Set the weights of PE3, PE6 and PE9 on the profiled device to $(w_{13}, w_{23}, w_{33})$. Set other weights to zero. Collect a phase-1 template trace for each input of the original set of traces.
    \item Subtract phase-1 template traces from the corresponding original traces. Perform DPA on column 2 (PE2, PE5, PE8) using updated traces. Run DPA for 1D array. Retrieve weights $(w_{12}, w_{22}, w_{32})$. 
    \item Set the weights of PE3, PE6 and PE9 on the profiled device to $(w_{13}, w_{23}, w_{33})$.
    Set the weights of PE2, PE5 and PE8 to $(w_{12}, w_{22}, w_{32})$. Set other weights to zero. Collect a phase-2 template trace for each input of the original set of traces.
    \item Subtract phase-2 template traces from the corresponding original traces. Perform DPA on column 1 (PE1, PE4, PE7) using updated traces. Run DPA for 1D array. Retrieve the final set of weights $(w_{11}, w_{21}, w_{31})$.
\end{itemize}

\begin{figure*}
\begin{minipage}[b]{0.325\linewidth}
\includegraphics[width=\linewidth]{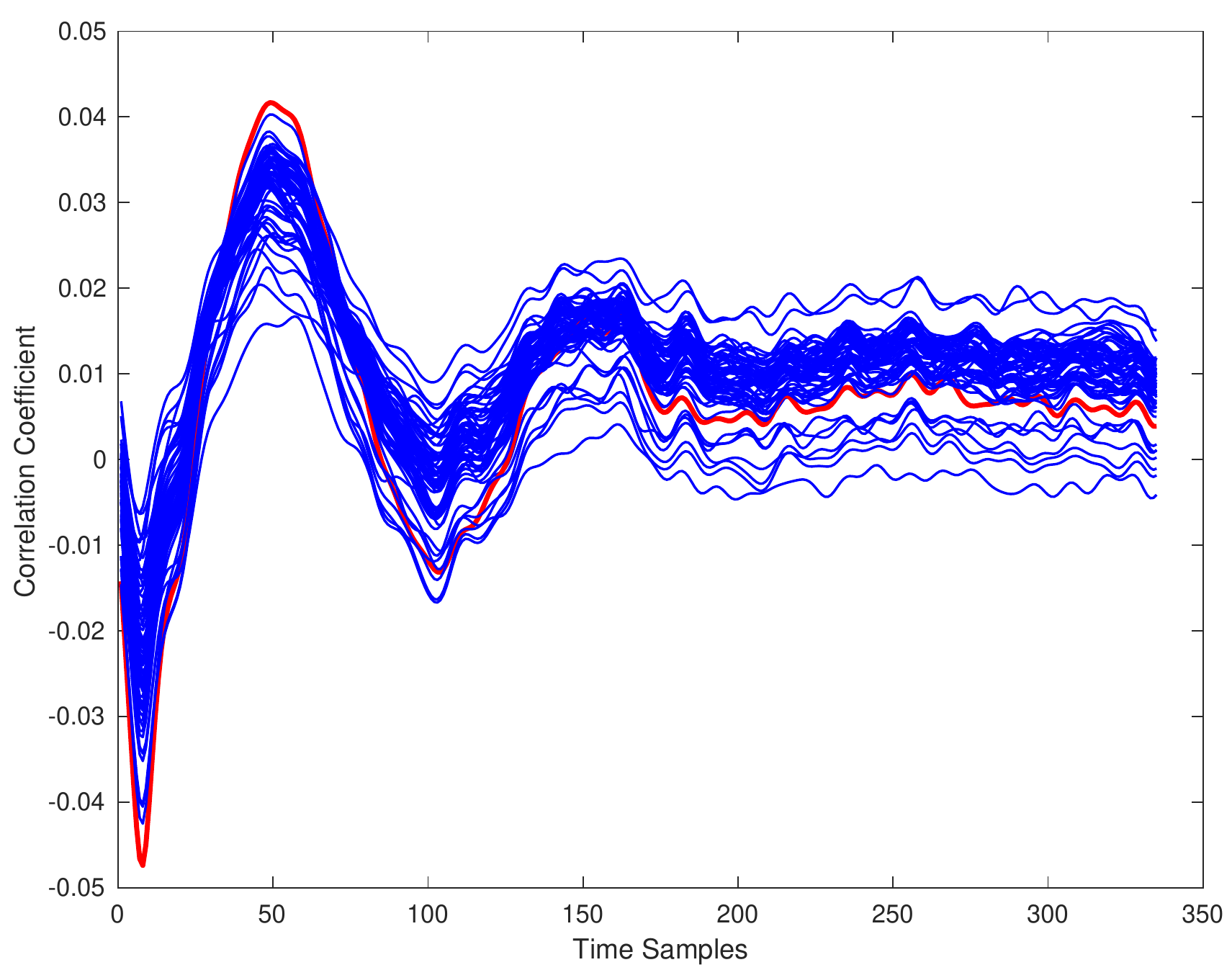}\\
\subcaption{DPA on $(w_{11}, w_{21}, w_{31})$.}\label{dpa_mv_col1}
\end{minipage}
\begin{minipage}[b]{0.325\linewidth}
\includegraphics[width=\linewidth]{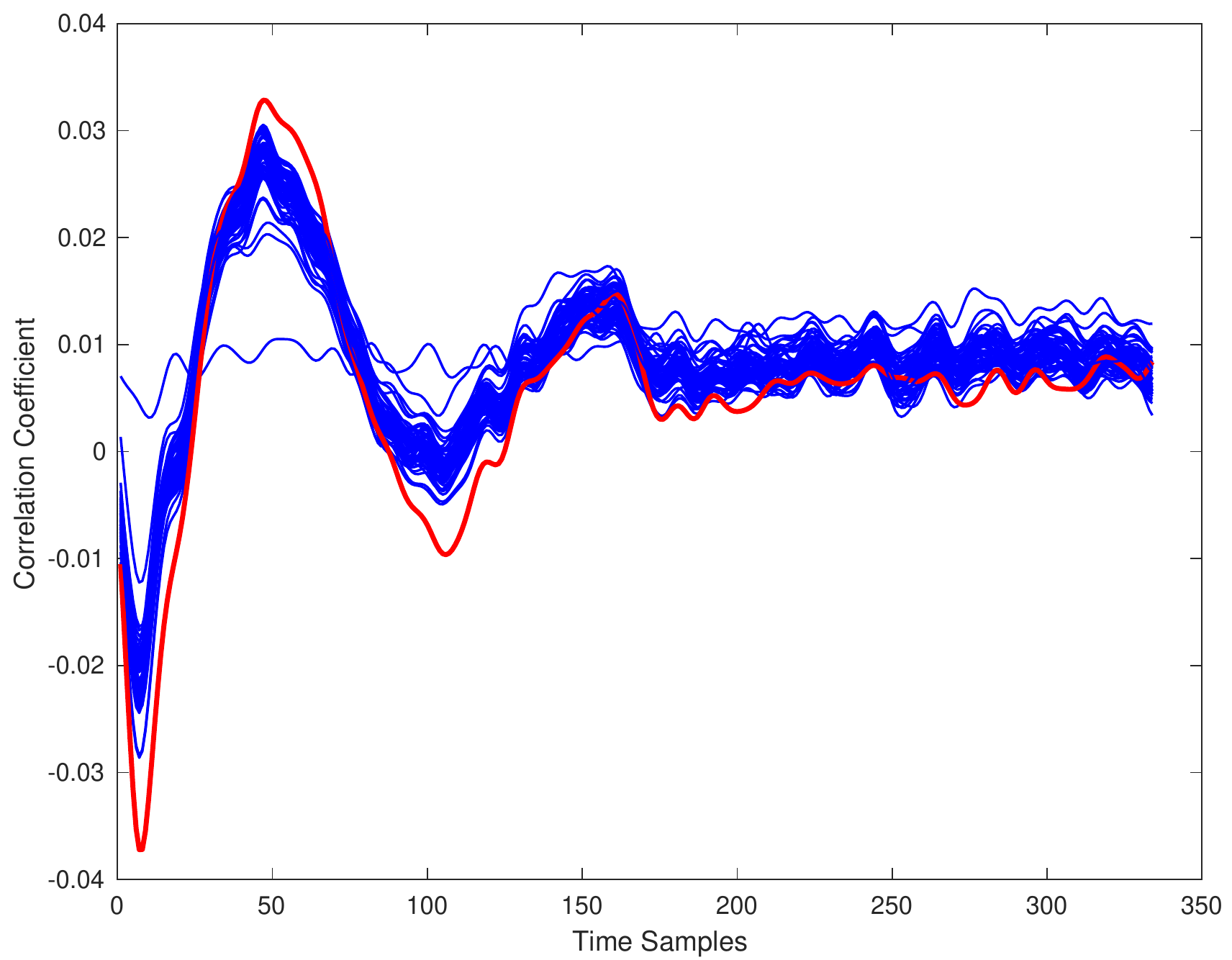}\\
\subcaption{DPA on $(w_{12}, w_{22}, w_{32})$.}\label{dpa_mv_col2}
\end{minipage}
\begin{minipage}[b]{0.325\linewidth}
\includegraphics[width=\linewidth]{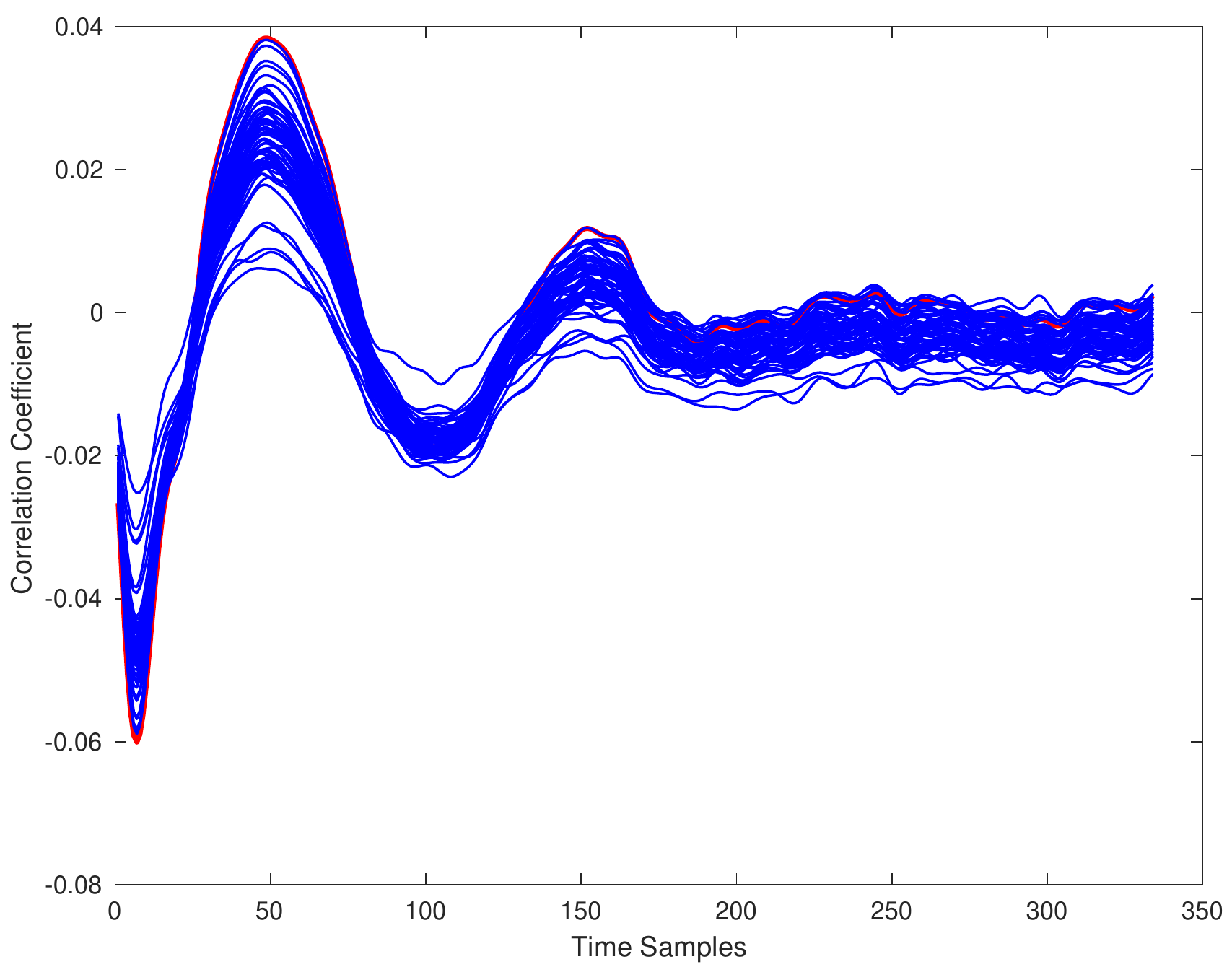}\\
\subcaption{DPA on $(w_{13}, w_{23}, w_{33})$.}\label{dpa_mv_col3}
\end{minipage}
\caption{Multi-phase template-based DPA results with 60K traces. The attack succeeds to fully retrieve the weights on matrix-vector multiplication accelerator. Red curves represent the correct combination.}
\label{dpa_mv_all}
\end{figure*}

We collect 60K traces from the matrix-vector multiplication accelerator and perform the above attack. The process requires 120K template traces to be collected in total. The attack succeeds in retrieving \emph{all} weights from the 2D array, Fig.\ref{dpa_mv_all}. We summarize the results for each PE column in Table \ref{tab:dpa_mv_results_multi_phase_profile}, identifying the MTD for each column.

\begin{table}[htbp]
    \centering
    \begin{tabular}{|c|c|c|c|c|}
    \hline
         Weight & Correct Guess & Rank & Correlation & MTD  \\
    \hline
         $(w_{11}, w_{21}, w_{31})$ & (23, -107, 74) & 1 & 0.0588 & 20000 \\ 
    \hline
         $(w_{12}, w_{22}, w_{32})$ & (120, 73, -96) & 1 & 0.0398 & 40000 \\ 
    \hline 
         $(w_{13}, w_{23}, w_{33})$ & (-6, -31, 17) & 1 & 0.0755 & 20000 \\ 
    \hline
    \end{tabular}
    \caption{Results of multi-phase template-based DPA for matrix-vector multiplication accelerator. The attack retrieves all weights from the accelerator with a MTD of 40K traces.}
    \label{tab:dpa_mv_results_multi_phase_profile}
\end{table}

We now analyze the time complexity of the multi-phase template-based DPA. The attack needs to apply a 1D DPA for each profiling phase of an individual PE column. Therefore, the complexity of the attack is proportional to the number of PEs in the 2D array. The cost of template construction is proportional to the number of PE columns. The complexity  also depends on the number of guesses to record for the first weight of each PE column.

\section{Discussion}

The time complexity of the DPA-based model extraction attack scales up linearly with the number of weights. Each weight vector loaded onto the systolic array is retrieved individually. To retrieve a large DNN model, the cost scales up with the model size. The linear increase of cost due to model size is inevitable in both side-channel-based \cite{csi_nn, masked_net, cema_single_pe} and query-based \cite{high_fidelity_extraction, cryptanalytic_extraction_dnn} model extraction attacks. Techniques to reduce MTD of each individual weight vector need to be investigated for a higher attack efficiency. 

Attacks based on EM measurements have the potential to further improve attack efficiency. EM attacks allow to localize leakage from individual components of the circuit, which can improve the SNR of the collected traces significantly. We anticipate that EM attacks with a high resolution EM probe, that is able to localize leakage from individual PE columns or even PEs, can break the 1D/2D arrays faster.

In this work, spatial DNN accelerators are shown to be vulnerable to DPA-based model extraction attacks. It is useful to consider some countermeasures to reduce or even eliminate side-channel leakage. Various countermeasures for embedded cryptographic hardware have been demonstrated over the years, such as masking \cite{masking}, \cite{higher_order_masking}, which adds random values to intermediate computations, and hiding \cite{wddl_prototype_ic}, \cite{equal_switch_cap}, \cite{ro_noise}, which aims to hide the power draw of the actual computation. These techniques are likely to also work on DNN accelerators. For instance, the masking scheme could be adopted to obfuscate the intermediate partial sums generated by MAC operations, which would break the correlation between the power models and the actual power consumption. Hiding could be realized by using dual-rail logic, or adding noise to hide the actual MAC power.

\section{Conclusion}

We investigate the vulnerability of spatial DNN accelerators using a general 8-bit number representation to DPA-style attacks. Specifically, we investigate two systolic array architectures based on the weight-stationary dataflow: (1) a 3 $\times$ 1 array for a dot-product operation, and (2) a 3 $\times$ 3 array for matrix-vector multiplication.
Both are implemented on the SAKURA-G FPGA board. We show that both architectures are ultimately vulnerable. A conventional DPA succeeds fully on the 1D array, requiring 20K power measurements. However, the 2D array exhibits higher security even with 460K traces. We show that this is because the 2D array intrinsically entails multiple MACs simultaneously dependent on the same input. However, we find that a novel template-based DPA with multiple profiling phases is able to fully break the 2D array with only 40K traces. Novel countermeasures need to be investigated to protect spatial DNN accelerators from such attacks.

\section{Acknowledgements}
This research was made possible by the support from the National Science Foundation under award 1901446. We also thank the anonymous reviewers for their feedback.

\bibliographystyle{ACM-Reference-Format}
\bibliography{main}

\end{document}